\documentclass[fleqn,12pt,twoside]{article}
\usepackage{espcrc1}

\usepackage{amssymb}
\usepackage[]{graphicx}

\def\pzp{p_0^{\prime}}
\def\pmp{p^{\prime}}

\def\eb{e_{\beta}}

\def\dbo{\partial_{\beta_{1}^2}}
\def\dbt{\partial_{\beta_{2}^2}}
\def\db{\partial_{\beta^2}}
\def\dbt{\partial_{\beta_2^2}}

\def\pp{{\bar p}}

\def\im{{\rm Im\,}}

\def\lc{L}
\def\lcp{{L}^{\prime}}
\def\lcpp{{L}^{\prime\prime}}

\title{Separable Kernel of Nucleon-Nucleon Interaction
in the Bethe-Salpeter Approach for $J=0,1$
\thanks{The work was supported in part by the RFBR grants {\em N} 00-15-96737 and
{\em N} 02-02-16542.}}

\author{S.G.\ Bondarenko\address[JINR]
{Joint Institute for Nuclear Research, Dubna 141980, Russia}
\thanks{bondarenko@jinr.ru},
V.V.\ Burov\addressmark[JINR]
\address[RCNP]{Research Center for Nuclear Physics,
Osaka University, Osaka, 567-0047, Japan}, N.\
Hamamoto\addressmark[RCNP],
A.\ Hosaka\addressmark[RCNP],
Y.\ Manabe\addressmark[RCNP],
H.\ Toki\addressmark[RCNP]}

\begin{document}
\maketitle

\begin{abstract}
The solution for the nucleon-nucleon ($NN$)
$T$ matrix in the framework of the covariant Bethe-Salpeter
approach for a two spin-one-half particle system
with a separable kernel of interaction is analyzed.
The explicit analytical connection between
parameters of the separable kernel and low energy scattering
parameters, deuteron binding energy and phase
shifts is established.
Covariant separable kernels for positive-energy partial channels
with total angular momentum
$J=0$ ($^1S_0^{+}$, $^3P_0^{+}$) and $J=1$ ($^3S_1^{+}-^3D_1^{+}$,
$^1P_1^{+}$, $^3P_1^{+}$) are constructed by
using obtained relations.
\end{abstract}
\newline
\noindent
PACS number(s): 11.10.St, 13.75.Cs, 21.45.+v

\section{Formalism}
We start with the partial-wave decomposed Bethe-Salpeter equation
for the nucleon-nucleon $T$ matrix (in the rest frame of two-nucleon system):
\begin{eqnarray}
t_{\lcp\lc}(\pzp, \pmp, p_0, p; s) &=&
v_{\lcp\lc}(\pzp, \pmp, p_0, p; s)
\label{t01}\\
&+& \frac{i}{4\pi^3} \sum_{\lcpp} \int\, dk_0\,\int\,
k^2\, dk\,
\frac{v_{\lcp\lcpp}(\pzp, \pmp, k_0, k; s)
t_{\lcpp\lc}(k_0, k, p_0, p; s)}{(\sqrt{s}/2-e_k+i0)^2-k_0^2}.
\nonumber\end{eqnarray}
Here $t$ is the partial-wave decomposed $T$ matrix and
$v$ is the kernel of the $NN$ interaction, $e_k=\sqrt{k^2+m^2}$.
There is only one term in the sum for the singlet (uncoupled triplet) case
($L=J$) and there are two terms for the coupled triplet case ($L=J\mp1$).
We introduced square of the total momentum $s=P^2=(p_1+p_2)^2$ and the relative
momentum $p=(p_1-p_2)/2$ [$p^{\prime}=(p^{\prime}_1-p^{\prime}_2)/2$] (for details,
see reference~\cite{report}).

Assuming the separable form (rank I) for the partial-wave decomposed
kernels of $NN$ interactions
\begin{eqnarray}
v_{\lcp\lc}(\pzp, \pmp, p_0, p; s) = \lambda g^{[\lcp]}(\pzp, \pmp) g^{[\lc]}(p_0, p),
\label{t04}\end{eqnarray}
we can solve eq.~(\ref{t01}) and write for the $T$ matrix:
\begin{eqnarray}
t_{\lcp\lc}(\pzp, \pmp, p_0, p; s) = \tau(s) g^{[\lcp]}(\pzp, \pmp) g^{[\lc]}(p_0, p),
\label{t05}\end{eqnarray}
with the function $\tau(s)$ being:
\begin{eqnarray}
\tau(s) = 1/(\lambda^{-1} + h(s)).
\label{t06}\end{eqnarray}
Function $h(s)$ has the following form:
\begin{eqnarray}
h(s) = \sum_{\lc} h_{\lc}(s)= -\frac{i}{4\pi^3}\, \int\, dk_0\,\int\, k^2\, dk\,
\sum_{\lc} \frac{[g^{[\lc]}(k_0,k)]^2}{(\sqrt{s}/2-e_k+i0)^2-k_0^2}.
\label{t07}\end{eqnarray}

We use following normalization condition for the on-mass-shell $T$ matrix
for the singlet (uncoupled triplet) case:
\begin{eqnarray}
t(s) \equiv t(0,\pp,0,\pp,s) = - \frac{16 \pi}{\sqrt{s}\sqrt{s-4m^2}}\,
e^{i\delta}\, \sin{\delta},
\label{t03}\end{eqnarray}
and for the coupled triplet case:
\begin{eqnarray}
t(s) =
\frac{8\pi i}{\sqrt{s}\sqrt{s-4m^2}}
\left(
\begin{array}{cc}
\cos{2\epsilon}\ e^{2i\delta_{<}} - 1 &
i\sin{2\epsilon}\ e^{i(\delta_{<}+\delta_{>})} \\
i\sin{2\epsilon}\ e^{i(\delta_{<}+\delta_{>})} &
\cos{2\epsilon}\ e^{2i\delta_{>}} - 1\\
\end{array}
\right),
\label{t10}\end{eqnarray}
with $\pp = \sqrt{s/4-m^2} = \sqrt{2mT_{lab}}$. We introduced phase shifts
$\delta\equiv\delta_{L=J}$, $\delta_{\lessgtr}\equiv\delta_{L=J\mp 1}$ and
mixing parameter $\epsilon$.

Bound state, if exist, is described by simple pole in the $T$ matrix.
Using eq.~(\ref{t06}) we can write
($M_b=2m-E_b$, $E_b$ is the energy of bound state):
\begin{eqnarray}
\lambda^{-1} = - h(s=M_b^2).
\label{t10a}\end{eqnarray}

We introduce also the low-energy parameters -- scattering length $a_{\lc}$ and
effective range $r_{\lc}$ -- by the following equation:
\begin{eqnarray}
\pp^{2\lc+1}\, {\cot}\, \delta_{\lc}(s) =
- 1/a_{\lc} + \frac{r_{\lc}}{2}\pp^2 + {\cal O}(\pp^3)
\label{t03a}\end{eqnarray}

At this point by using eq.~(\ref{t05}) and calculating $T$ matrix
on the mass-shell ($p_0=p_0^{\prime}=0,p=p^{\prime}=\pp$) we can connect
internal parameters of the $NN$ kernel and observables -- phase shifts, bound state
energy and low-energy parameters.

We use covariant generalization of the {\em Yamaguchi}~\cite{yam}
functions for $g^{[\lc]}(k_0,k)$:
\begin{eqnarray}
g^{[S]}(k_0,k) = \frac{1}{k_0^2-k^2-\beta_0^2+i0},
\label{t11a}
\end{eqnarray}
\vskip -7mm
\begin{eqnarray}
g^{[P]}(k_0,k) = \frac{\sqrt{|-k_0^2+k^2|}}{(k_0^2-k^2-\beta_1^2+i0)^2},
\end{eqnarray}
\vskip -7mm
\begin{eqnarray}
g^{[D]}(k_0,k) = \frac{C(k_0^2-k^2)}{(k_0^2-k^2-\beta_2^2+i0)^2}.
\label{t11}\end{eqnarray}

Let us consider function $h_0(s)$ defined in the following way
(we introduce explicit dependence on $\beta$):
\begin{eqnarray}
h_0(s,\beta) = -\,\frac{i}{4\pi^3}\, \db\, \int dk_0\,\int\, k^2\, dk\
\frac{1}{(\sqrt{s}/2-e_k+i0)^2-k_0^2}\
\frac{1}{k_0^2-\eb^2+i0},\label{t12}
\end{eqnarray}
here $\eb=\sqrt{k^2+\beta^2}$ and $\db = \partial/\partial\beta^2$.

Analyzing the analytic structure and properties of
the function $h_0(s,\beta)$ on $s$
we can rewrite it in the dispersion form:
\begin{eqnarray}
h_0(s,\beta) = \int\limits_{4m^2}^{+\infty}
\frac{\rho(s^{\prime},\beta)\,ds^{\prime}}{s^{\prime}-s-i\epsilon},
\label{t24a}
\end{eqnarray}
\vskip -7mm
\begin{eqnarray}
\rho(s^{\prime},\beta) = \theta (s^{\prime}-4m^2)
\rho_{\rm el}(s^{\prime},\beta) + \theta
(s^{\prime}-4(m+\beta)^2)\rho_{\rm in}(s^{\prime},\beta),
\nonumber\end{eqnarray}
with two spectral functions $\rho_{\rm el;in}$
({\em el} stands for {\em elastic} and {\em in} for {\em
inelastic}) which are connected with the imaginary parts as
\begin{eqnarray}
\rho(s^{\prime},\beta) = \frac{1}{\pi} \im h_0(s^{\prime},\beta) =
\frac{1}{2\pi i} (h_0-h_0^{*}).
\label{t25a}\end{eqnarray}

Calculating integral~(\ref{t24a}) we can obtain analytic expressions
for function $h_0(s,\beta)$ and taking into account
definitions~(\ref{t11a}-\ref{t11}) can connect
all other functions $h^{[\lc]}$ with $h_0$ as:
\begin{eqnarray}
h^{[S]}(s,\beta_0) = h_0(s,\beta_0),
\end{eqnarray}
\vskip -7mm
\begin{eqnarray}
h^{[P]}(s,\beta_1) = -\frac{1}{2}\left[
\dbo+\frac{1}{3}\beta_1^2\dbo^2\right] h_0(s,\beta_1),
\end{eqnarray}
\vskip -7mm
\begin{eqnarray}
h^{[D]}(s,\beta_2) = C^2\left[
1+\beta_2^2\dbt+\frac{1}{6}\beta_2^4\dbt^2\right] h_0(s,\beta_2).
\label{t14}
\end{eqnarray}

\section{Calculations and results}

We can now calculate internal parameters of the $NN$ kernel
by using obtained above equations to reproduce experimental
values for the phase shifts (data are taken using SAID program
http://gwdac.phys.gwu.edu/),
deuteron energy and quadrupole moment,
low-energy parameters (data are from ref.~\cite{data}).

1. To find parameters $\lambda$ and $\beta$ in $^{1}S_0^+$ channel
we solve system of the nonlinear equations (exp stands for experimental,
$s$ - for singlet):
\begin{eqnarray}
a_s^{\rm exp} = a_s(\lambda,\beta),\qquad r_s^{\rm exp} = r_s(\lambda,\beta).
\end{eqnarray}

2. To find parameters $\lambda$, $\beta_0$, $\beta_2$ and $C$
in $^{3}S_1^+-^{3}D_1^+$ coupled channel we solve system
of the nonlinear equations ($t$ stands for triplet):
\begin{eqnarray}
a_t^{\rm exp} = a_t(\lambda,\beta_0,\beta_2,C),\qquad
E_d^{\rm exp} = r_0(\lambda,\beta_0,\beta_2,C),
\end{eqnarray}
\vskip -10mm
\begin{eqnarray}
p_d = p_d(\lambda,\beta_0,\beta_2,C),\qquad
q_d^{\rm exp} = q_d(\lambda,\beta_0,\beta_2,C).
\nonumber
\end{eqnarray}
Here we introduced $D$-wave pseudoprobability $p_d$.

3. To find parameters $\lambda$ and $\beta$ in uncoupled $^{3}P_0^+$, $^{1}P_1^+$
and $^{3}P_1^+$ channels we use procedure to minimize function:
\begin{eqnarray}
\chi^2 = \sum\limits_{i=1}^{n}
(\delta^{\rm exp}(s_i)-\delta(s_i))^2/(\Delta\delta^{\rm exp}(s_i))^2,
\end{eqnarray}
where $n$ is the number of the experimental points taking into account.

The results of calculations are given in tables~1 and~2 and
figs.~1-4.

\newpage

\noindent
{\small Table 1. Parameters for $^{1}S_0^+$ and $^{3}S_1^+-^{3}D_1^+$ channels.}
\\
\begin{tabular}{lcccc}
\hline
&$^1S_0^+$ & $^3S_1^+-^3D_1^+$ & $^3S_1^+-^3D_1^+$ &
$^3S_1^+-^3D_1^+$ \\
&          & ($p_d = 4\%$) & ($p_d = 5\%$) & ($p_d = 6\%$)\\
\hline
$\lambda$ (GeV$^2$) & -0.294254 & -0.499045 & -0.425235 & -0.359856\\
$\beta_0$ (GeV) & 0.224129 & 0.251248 & 0.246713 & 0.242291\\
$\beta_2$ (GeV) && 0.294096 & 0.324494 & 0.350217\\
C && 1.6489 & 2.4109 & 3.2801\\
\hline
\end{tabular}

\noindent
{\small Table 2. Parameters for $^{3}P_0^+$, $^{1}P_1^+$
and $^{3}P_1^+$ channels.}
\\
\begin{tabular}{lcccccc}
\hline
&$^3P_0^+$ & $^3P_0^+$ & $^1P_1^+$ & $^1P_1^+$ & $^3P_1^+$ & $^3P_1^+$\\
& n=3  & n=5 & n=4  & n=5 & n=7 & n=9\\
\hline
$\lambda$ (GeV$^2$) & -0.0295720 & -0.0161201 & 0.0915850 & 0.195296 &
0.312975 & 0.658122 \\
$\beta_1$ (GeV) & 0.238515 & 0.2186083 & 0.276724 & 0.308907 & 0.338898 &
0.381901 \\
\hline
\end{tabular}

\vskip 10mm

\noindent
\begin{minipage}{0.48\textwidth}
\includegraphics[width=\textwidth]{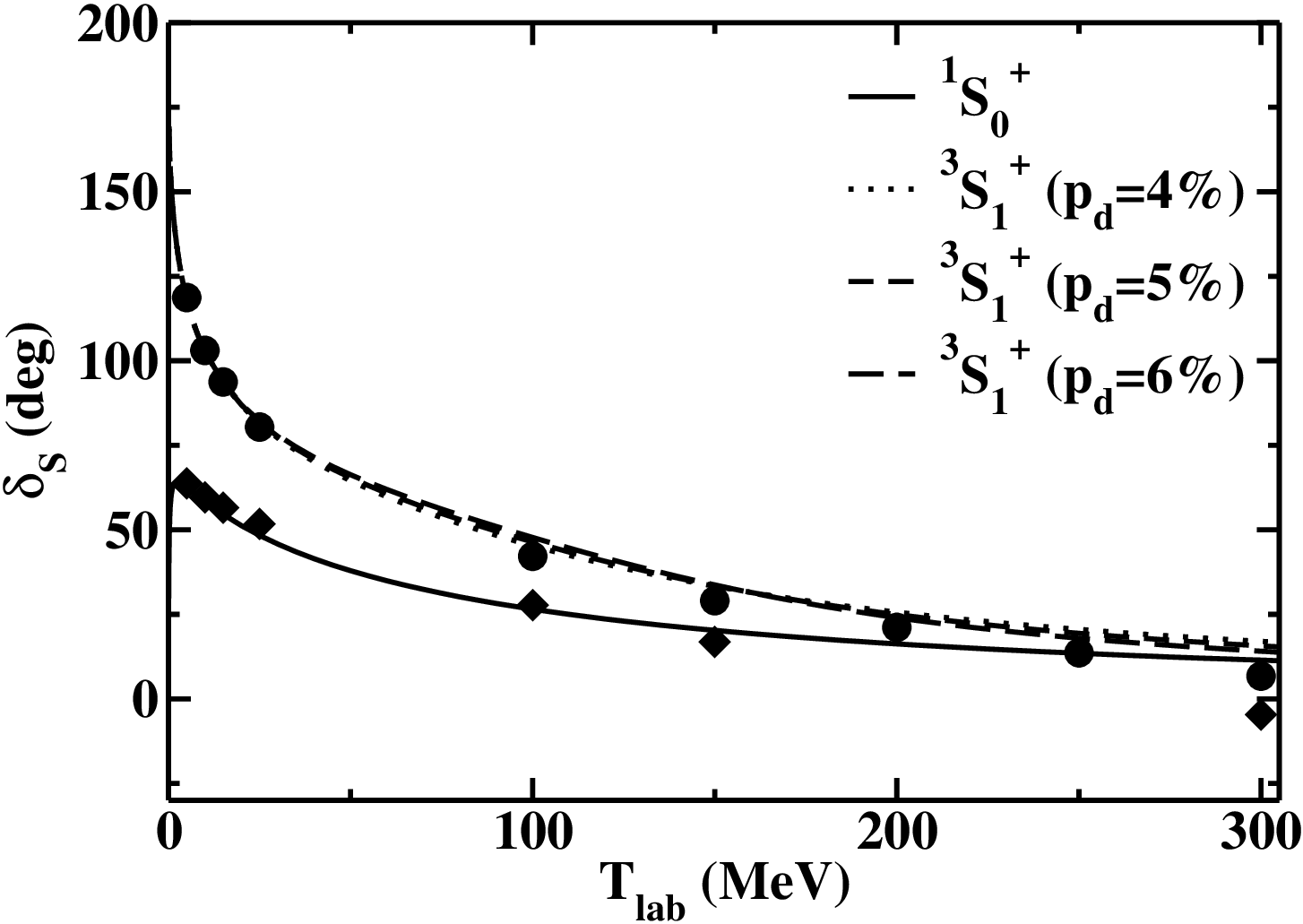}
\\[-1mm]
{\small Figure 1. $^1S_0^+$ and $^3S_1^+$ channels phase shifts.}
\end{minipage}
\hskip 3mm
\begin{minipage}{0.48\textwidth}
\includegraphics[width=\textwidth]{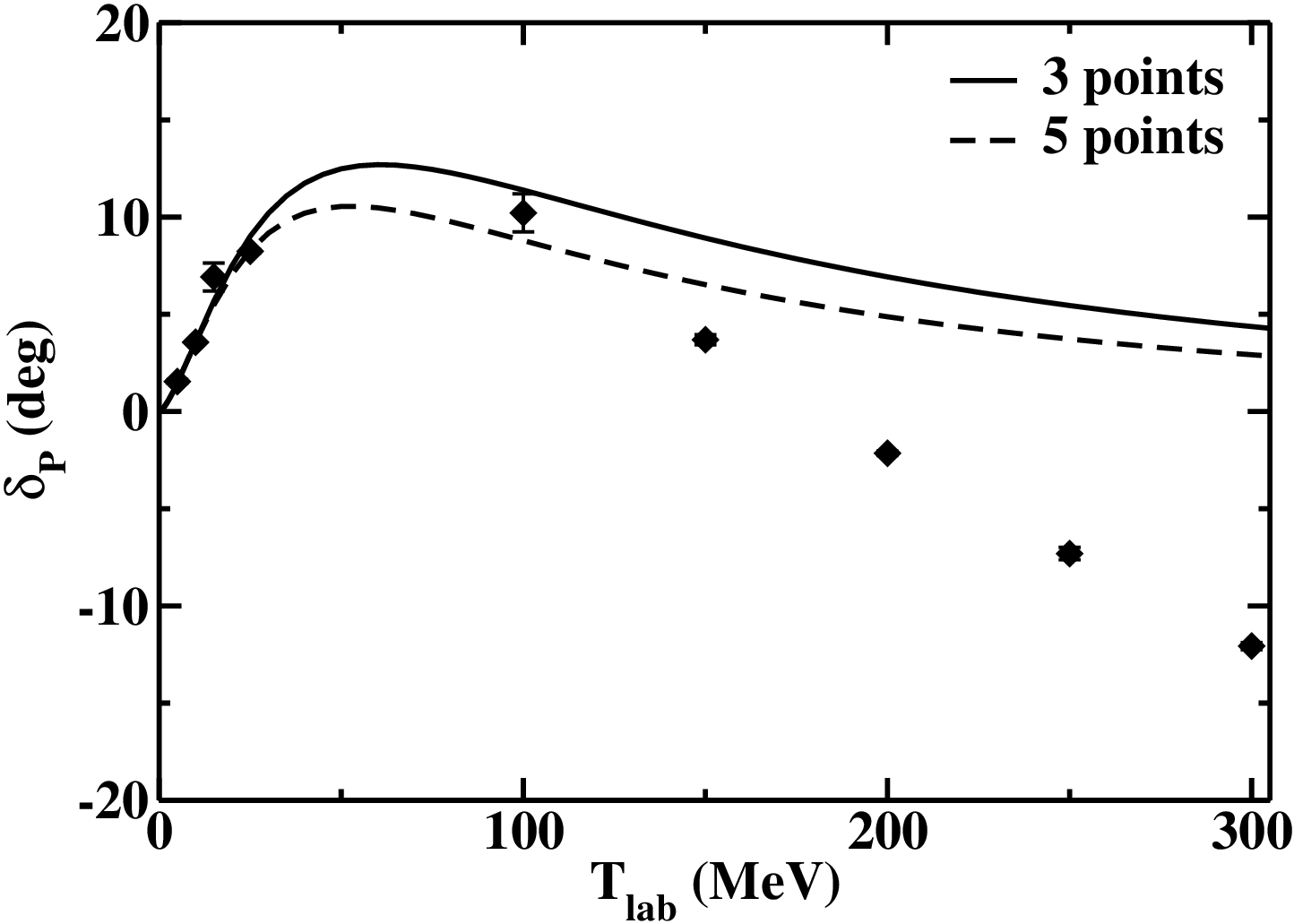}
\\[-1mm]
\phantom{aaa}\hskip 5mm {\small Figure 2. $^3P_0^+$ channel phase shifts.}
\end{minipage}

\vskip 10mm

\noindent
\begin{minipage}{0.48\textwidth}
\includegraphics[width=\textwidth]{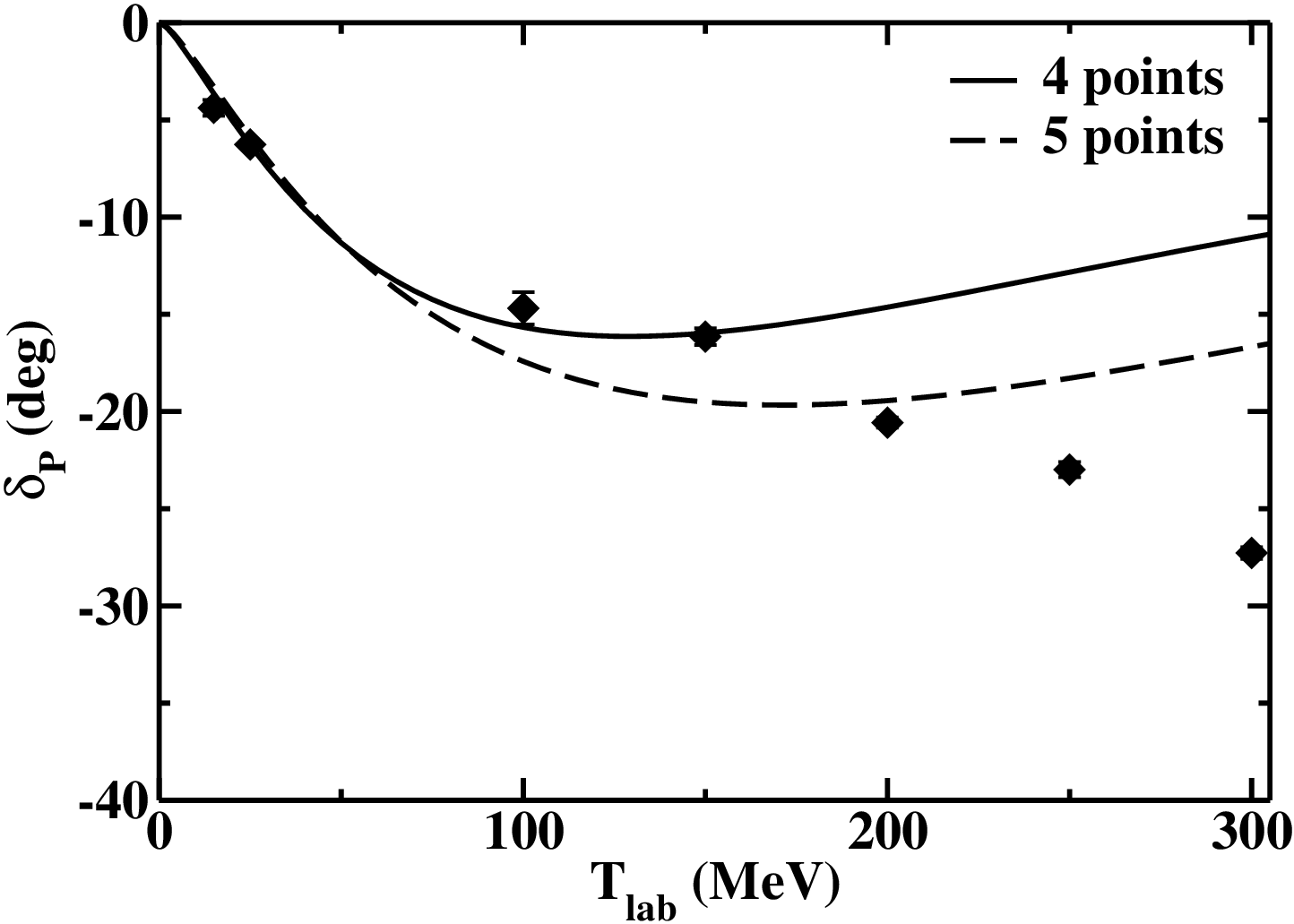}
\\[-1.5mm]
\phantom{aaa}\hskip 5mm {\small Figure 3. $^1P_1^+$ channel phase shifts.}
\end{minipage}
\hskip 3mm
\begin{minipage}{0.48\textwidth}
\includegraphics[width=\textwidth]{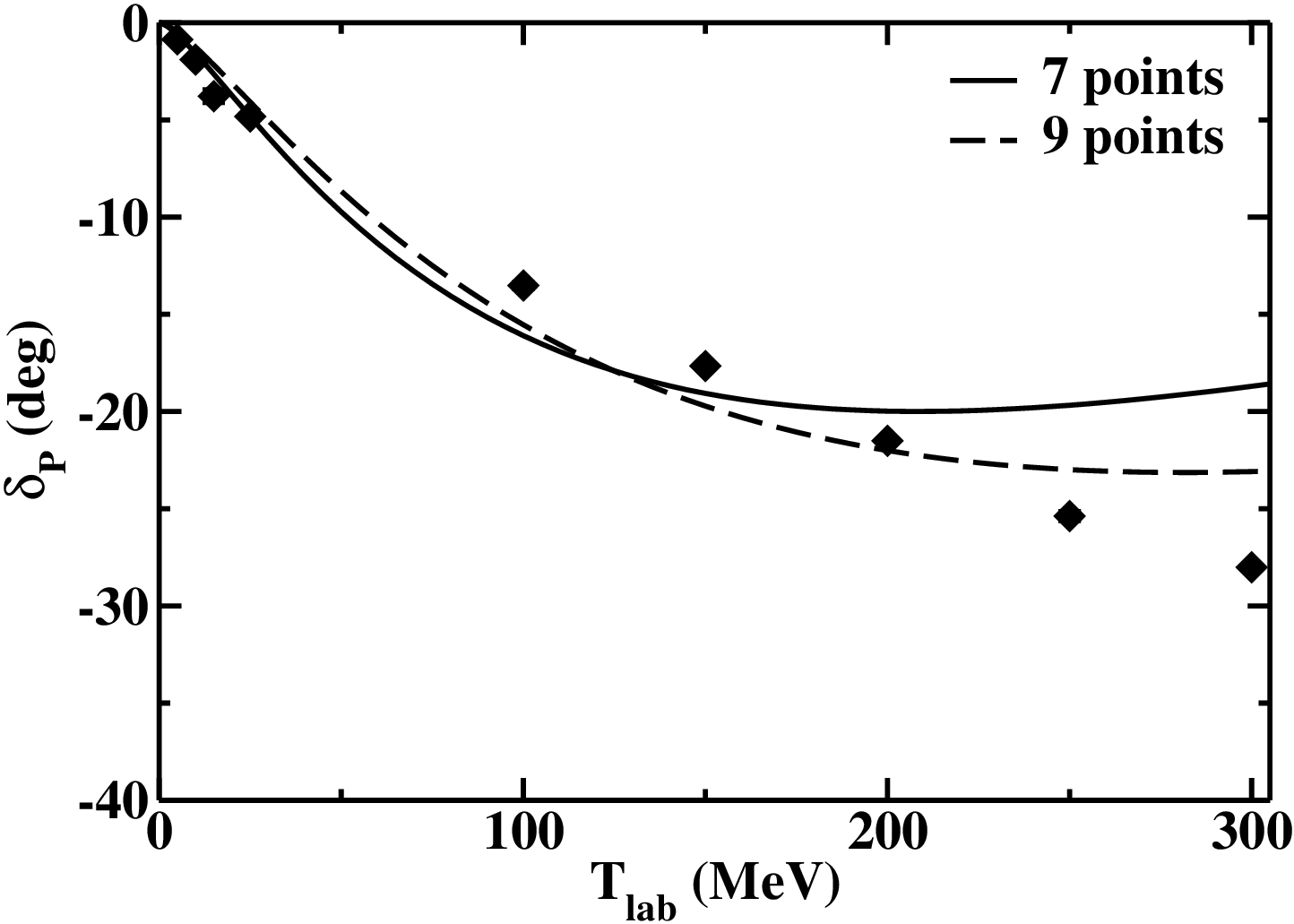}
\\[-1.5mm]
\phantom{aaa}\hskip 5mm {\small Figure 4. $^3P_1^+$ channel phase shifts.}
\end{minipage}


\begin{thebibliography}{}
\bibitem{report} S.G. Bondarenko, et al., Prog.Part.Nucl.Phys.,
{\bfseries 48} (2002) 449; nucl-th/0203069.
\bibitem{yam} Y. Yamaguchi, Phys. Rev. {\bfseries 95}, (1954) p.1628.
\bibitem{data} O. Dumbrajs {\em et al.},
Nucl. Phys. B {\bfseries 216} (1983) p.277.
\end{thebibliography}
\end{document}